\documentclass[a4paper,11pt]{article}
\usepackage{jinstpub} % for details on the use of the package, please
\usepackage{subcaption}
\usepackage{lineno}
\usepackage{siunitx}
\usepackage{comment}
\usepackage{amsmath}
\usepackage{upgreek}
\usepackage{textgreek}
\usepackage{enumitem}

\title{\boldmath Qualification of Bump Bonding in CMS Inner Tracker Pixel Modules for the Phase-2 Upgrade}

\author[a,1]{P. Assiouras,\note{Corresponding author.}}

\affiliation[a]{INFN Sezione di Firenze, Italy}
\emailAdd{panagiotis.assiouras@cern.ch}

\abstract{
	To fully exploit the increased luminosity of the HL-LHC, the CMS Inner Tracker is undergoing a major upgrade to withstand extreme radiation levels and data rates, while improving granularity and reducing material budget. The upgraded modules employ thin planar and 3D silicon pixel sensors, bump bonded to a new radiation-hard readout chip, the CROC, designed in 65 nm CMOS technology and powered via a serial scheme. Ensuring the quality of bump bonding between sensors and readout chips is critical for efficient detector operation. This work presents the qualification procedures developed to identify missing or defective bumps using multiple test methods, including crosstalk analysis, reverse/forward bias testing, and X-ray or beta source imaging. Results from prototype modules are presented and advantages and limitations of each method are discussed.
}


\arxivnumber{1234.56789} % only if you have one

\collaboration[c]{for the Tracker Group of the CMS Collaboration}

\proceeding{20$^{\text{th}}$ Anniversary Trento Workshop on Advanced Silicon Radiation Detectors\\
  4-6 Feb 2025\\
  FBK, Trento}

\begin{document}
	
\maketitle
\flushbottom

\section{Introduction}
\label{sec:intro}

The High-Luminosity LHC (HL-LHC) upgrade aims to increase the LHC's instantaneous luminosity up to 7.5$\times$ 10$^{34}$ cm$^{-2}$s$^{-1}$, enabling precision studies of the Higgs boson, rare processes, and searches beyond the Standard Model. This increase poses significant challenges for the detectors, which must operate under higher radiation levels and pileup rates of up to an average of 200 simultaneous collisions. To meet these conditions, all major LHC experiments--including the Compact Muon Solenoid (CMS \cite{CMS:2008xjf})--are undergoing a comprehensive upgrade (Phase-2 upgrade~\cite{CMS-TDR-014}).

\begin{figure}[!htbp] \centering \includegraphics[height=5.0cm,width=0.80\textwidth]{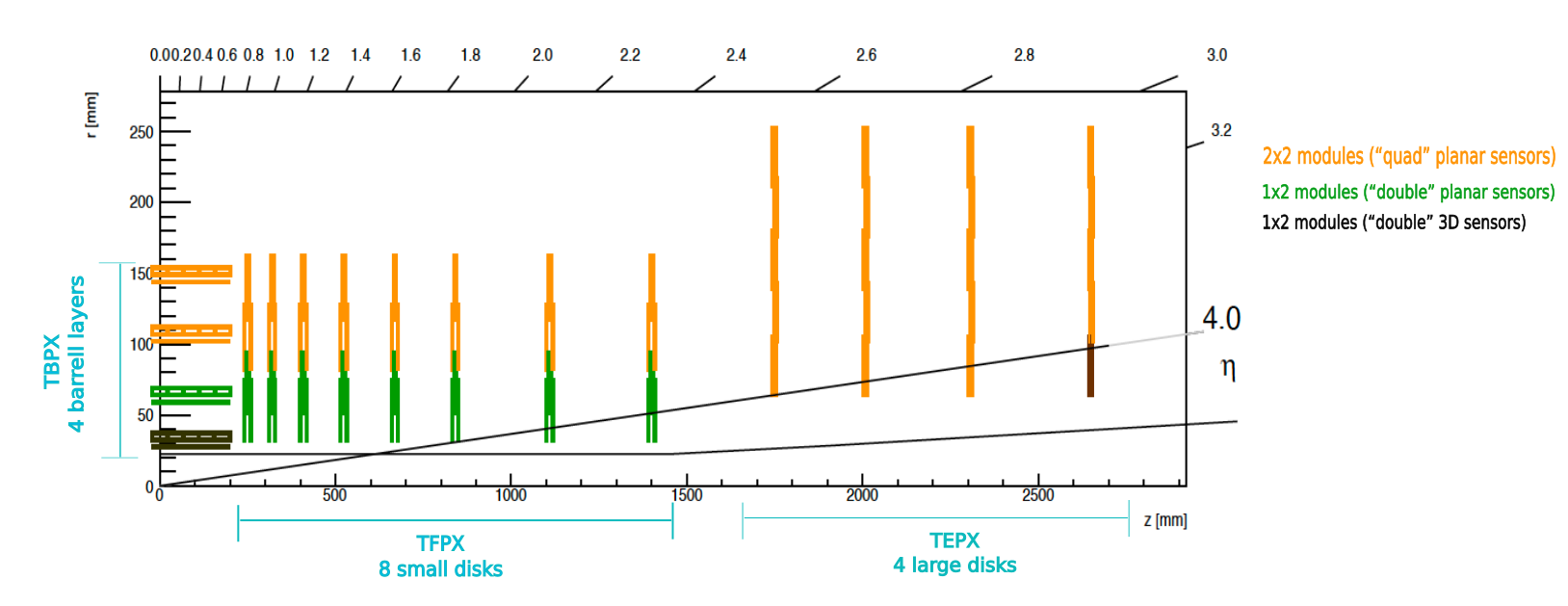} \caption{Schematic of one quarter of the CMS Inner Tracker layout in r-z view. The detector features four barrel layers (TBPX), 8 TFPX double-discs, and 4 TEPX double-discs per side.
	}
	\label{fig:cms_inner_tracker_phase2}
\end{figure}

The upgraded CMS Inner Tracker is designed to operate reliably under the high pileup and radiation conditions of the HL-LHC. The layout features three main subdetectors: the centrally positioned Barrel Pixel Detector (TBPX), and the Tracker Forward Pixel Detector (TFPX) and Tracker Endcap Pixel Detector (TEPX) in the forward regions, as shown in figure~\ref{fig:cms_inner_tracker_phase2}. This layout provides hermetic tracking coverage up to $|\upeta| \approx 4$ with high granularity and modularity. The innermost barrel layer (TBPX L1) will be subject to the most extreme radiation, with expected fluences reaching 2.6 $\cdot$ 10$^{16}$ $\text{n}_{\text{eq}}/\text{cm}^2$ and doses up to 1.3 Grad after an integrated luminosity of 3000 fb$^{-1}$. To ensure long-term performance under these conditions, the TBPX L1 modules are designed to be replaceable midway through HL-LHC operation.

There are two types of modules used in the Phase-2 upgrade of the CMS Inner Tracker. The two inner layers of the TBPX and the two inner rings of the TFPX will be populated with double-chip modules (shown in green in figure~\ref{fig:cms_inner_tracker_phase2}), while the two outer layers of the TBPX, along with the outer rings of both the TFPX and the TEPX, will use quad-chip modules (shown in yellow). Each module is built on a High-Density Interconnect (HDI) printed circuit board, which distributes power, data, and control signals to the readout chips and supports silicon pixel sensors that are 150 \si{\micro\meter} thick and fabricated on 6-inch wafers. Two sensor technologies are employed--planar and 3D--both featuring a 25$\times$100 \si{\micro\meter} pixel pitch to ensure high spatial resolution and radiation tolerance \cite{CMS-TDR-014}. These sensors are bump bonded to the CMS Readout Chip (CROC), a radiation-hard ASIC developed in 65 nm CMOS by the RD53 collaboration~\cite{Loddo2024}. The CROC offers low power consumption, high data throughput, and integrated Shunt-LDO regulators for efficient serial powering. After bump bonding, the resulting sensor-chip assemblies (bare modules) are sent to CMS assembly centers, where they are mounted onto the HDI boards with high mechanical and electrical precision.

%The upgraded CMS Inner Tracker is designed to operate reliably under the high pileup and radiation conditions of the HL-LHC. In the innermost barrel layer (TBPX L1), radiation levels are expected to reach up to 2.6 $\cdot$ 10$^{16}$ $\text{n}_{\text{eq}}/\text{cm}^2$ and 1.3 Grad after 3000 fb$^{-1}$. To meet these challenges, the layout features high segmentation and modularity (figure \ref{fig:cms_inner_tracker_phase2}). This includes the possibility of replacing TBPX L1 modules midway through HL-LHC operation to ensure continued performance. The Tracker Forward Pixel Detector (TFPX) and Tracker Endcap Pixel Detector (TEPX) comprise a series of double-discs per side, each containing concentric rings of pixel modules, extending the pseudorapidity coverage to $|\upeta| \approx 4$.

%There are two types of modules used in the Phase-2 upgrade of the CMS Inner Tracker. The two inner layers of the TBPX and the two inner rings of the TFPX will be  populated with double-chip modules (shown in green in figure~\ref{fig:cms_inner_tracker_phase2}), while the two outer layers of TBPX, along with the outer rings of both TFPX and the TEPX, will use quad-chip modules (shown in yellow). Each module is built on a High-Density Interconnect (HDI) printed circuit board, which distributes power, data, and control signals to the readout chips. 

A dedicated quality control framework has been established within CMS to ensure each module meets the stringent requirements for HL-LHC operation. A key focus is the evaluation of bump bonding between sensors and readout chips, as defective bumps degrade tracking performance. This paper outlines qualification procedures--crosstalk analysis, reverse/forward bias testing, and beta source imaging--used to identify such defects. Results from prototype module testing are presented, highlighting the strengths and limitations of each method.

\section{Module quality control}

\begin{figure}[!htbp]
	\centering
	\includegraphics[height=6.5cm,width=0.6\textwidth]{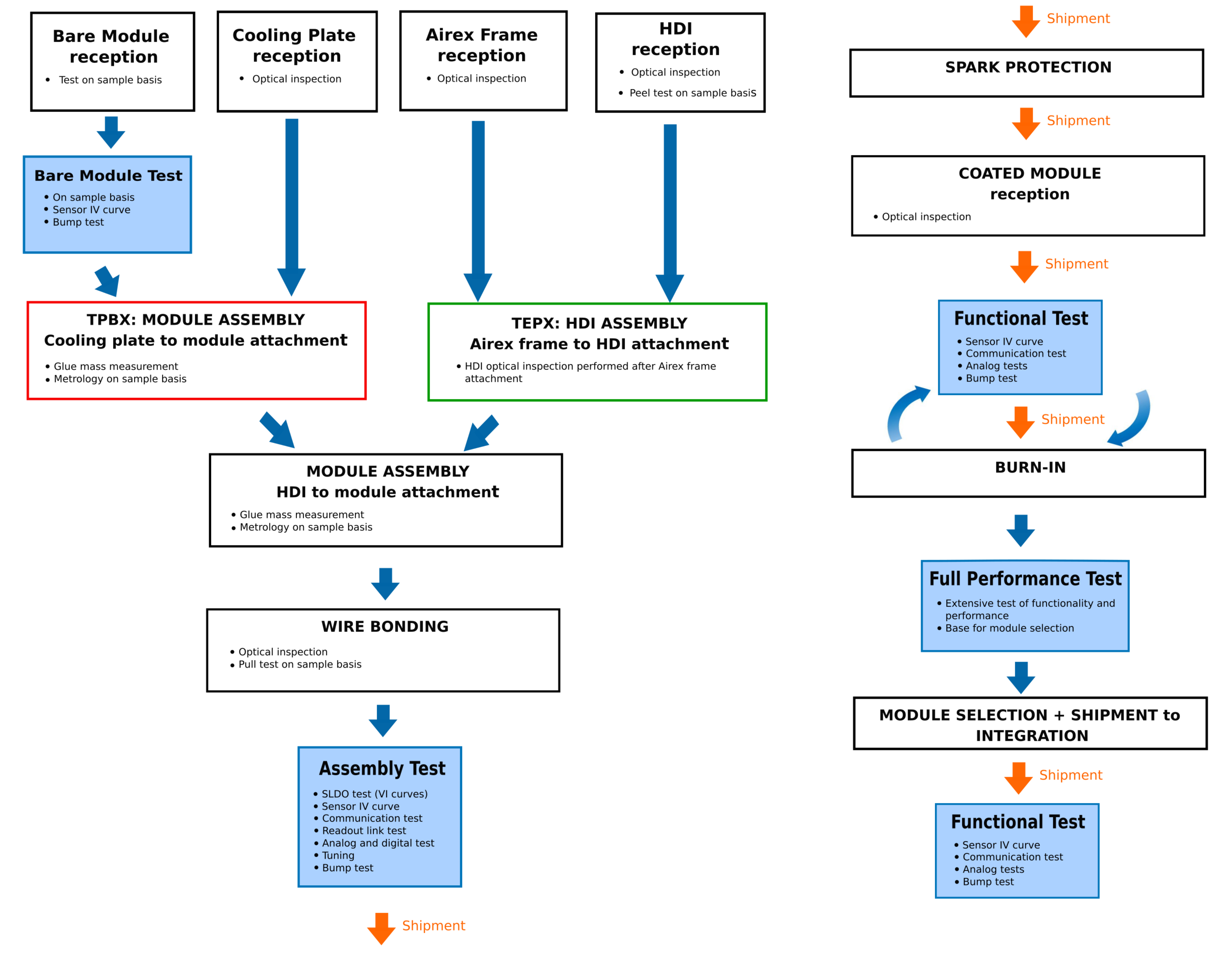}
	\caption{Module production and quality control workflow for CMS Inner Tracker pixel modules. The process begins with the reception and inspection of components such as bare modules, cooling plates, Airex frames, and HDIs. Assembly proceeds through multiple stages including module assembly, wire bonding, and spark protection. At each stage, specific quality control tests are conducted. Modules undergo repeated functional tests, after shipment and coating, followed by burn-in and a final Full Performance Test.
	\label{fig:QC_module_procedure}}
\end{figure}

During full-scale production, nearly 4000 pixel modules will be tested. This large number necessitates a stable, efficient, and streamlined measurement procedure, which is designed to be as automated as possible. At each production stage, modules undergo rigorous quality control, including checks for electrical functionality, structural integrity, and thermal stability. Figure \ref{fig:QC_module_procedure} illustrates the production flow and the corresponding quality control steps.

Initial testing is performed at the bare module level to assess the integrity of the bare-modules prior to assembly. This includes optical inspection of the bare modules, I-V measurements of the sensors, and low-voltage tests on the readout chips using a dedicated probe card, as described in \cite{GRIPPO2022167496}. Once assembled, each module undergoes the Assembly Test, which verifies functionality that may be affected by mechanical assembly steps or wire bonding. Modules that fail to meet the electrical requirements at this stage are excluded from further production steps. Throughout the production process, modules are shipped multiple times, particularly for spark protection and coating. After each shipment, a Functional Test is conducted--most importantly after spark protection and again upon reception at the integration site--to verify that shipping and processing have not affected the module's performance.

Final qualification is performed during the Full Performance Test, which provides the data used for module selection. Modules are tested both at 17 $^{\circ}$C and at the nominal operating temperature of -20 $^{\circ}$C. This test is performed after the burn-in stage, where modules undergo at least 10 thermal cycles between -35 $^{\circ}$C and +40 $^{\circ}$C in a dedicated cold box to ensure thermal robustness and long-term stability. Bump bond integrity is re-evaluated after thermal cycling, as the temperature-induced stress can expose weak or marginal bump connections.

\section{Bump bonding tests}

Bump bonding, also referred to as hybridization, is a critical process in the construction of bare modules, in which each pixel of the silicon sensor is electrically connected to its corresponding readout pixel on the CROC chip via micro-bumps. This operation demands sub-micron alignment accuracy, uniform bump formation, and reliable thermal compression to ensure high-quality connections. For the CMS Inner Tracker upgrade, hybridization is carried out by specialized industrial partners, including IZM, Advafab, and Micross. To ensure high bump yield throughout production, dedicated qualification procedures have been implemented.

\subsection{Crosstalk method}

The crosstalk method assesses the integrity of bump bonds between the sensor and the readout chip by utilizing the internal charge injection circuitry of the front-end electronics. A large signal (39,000 e$^-$) is injected into a specific pixel, and the response of neighboring pixels is measured. The pixel thresholds are initially tuned to approximately 1200 e$^-$, allowing the induced signals from capacitive coupling to be reliably detected. In the case of a properly connected bump bond, the injected charge induces measurable signals in adjacent pixels. Conversely, if the bump bond is missing, this coupling is suppressed, and neighboring pixels exhibit significantly reduced or no response. This technique is non-destructive, requires no external radiation sources, and can be easily automated, making it well-suited for use during module production.

To increase the method's sensitivity, two distinct pixel injection patterns are used: coupled and uncoupled. This distinction arises from the different layout geometries of the sensor and the readout chip. The silicon sensors feature rectangular pixels with a pitch of 25$\times$100 \si{\micro\meter^2}, while the CROC readout chip uses a 50$\times$50 \si{\micro\meter^2} bump pad pitch \cite{CMS-TDR-014,RD53CManual}. This layout leads to varying spatial relationships between injected and neighboring pixels, resulting in different levels of capacitive coupling. Using both injection patterns improves the robustness of the crosstalk method and helps avoid misclassification of pixels affected by neighboring defects. 
 
\begin{figure}[!htbp]
	\centering
	\begin{subfigure}{0.45\textwidth}
		\centering
		\includegraphics[height=5.0cm, width=\textwidth]{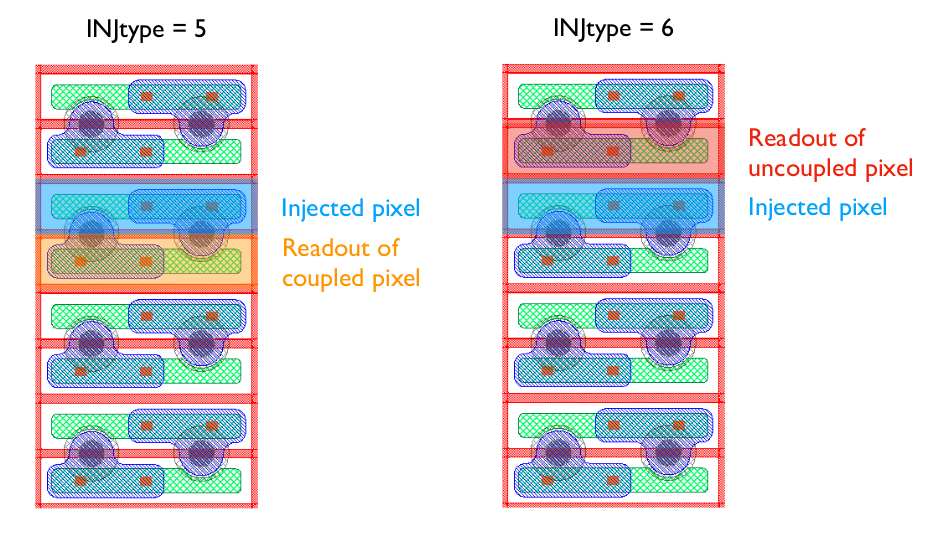} 
		\caption{\label{fig:coupled_and_uncoupled pixels}}
	\end{subfigure} \quad
	\begin{subfigure}{0.50\textwidth}
		\centering
		\includegraphics[height=5.0cm, width=0.9\textwidth]{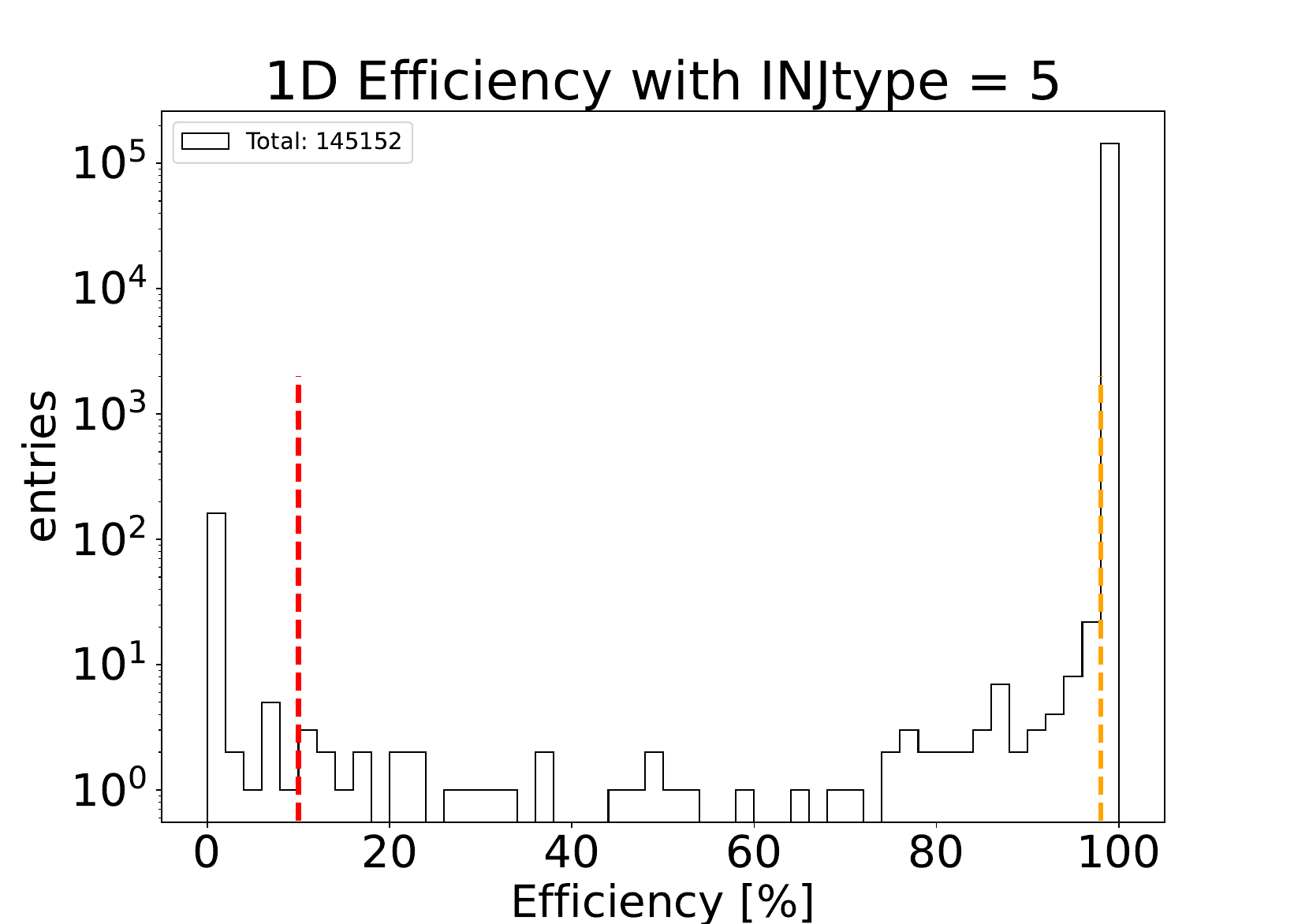}
		\caption{\label{fig:efficiency_map_classification}}
	\end{subfigure}
	\caption{Crosstalk-based bump bond classification. (a) Injection patterns: coupled and uncoupled. (b) Example histogram used to classify pixels as functional (>95\%), missing (<10\%), or problematic.
	}
\end{figure}

Figure \ref{fig:coupled_and_uncoupled pixels} illustrates these two configurations: the coupled case (left) shows stronger crosstalk in nearby pixels due to geometric proximity and enhanced capacitive coupling, while the uncoupled case (right) results in more isolated signal responses. The coupling strength is further influenced by the layout of the sensor's metal layers, as overlapping metal and implant structures increase inter-pixel capacitance. After injection, the efficiency of signal detection in surrounding pixels is evaluated, as shown in figure \ref{fig:efficiency_map_classification}. Pixels with efficiency below 10$\%$ for both injection configurations are flagged as missing, those above 95$\%$ are considered functional, and the remaining are labeled as problematic. These thresholds were optimized based on studies on prototype modules.

\begin{figure}[!htbp]
	\centering
	\begin{subfigure}{0.48\textwidth}
		\centering
		\includegraphics[height=5.0cm, width=\textwidth]{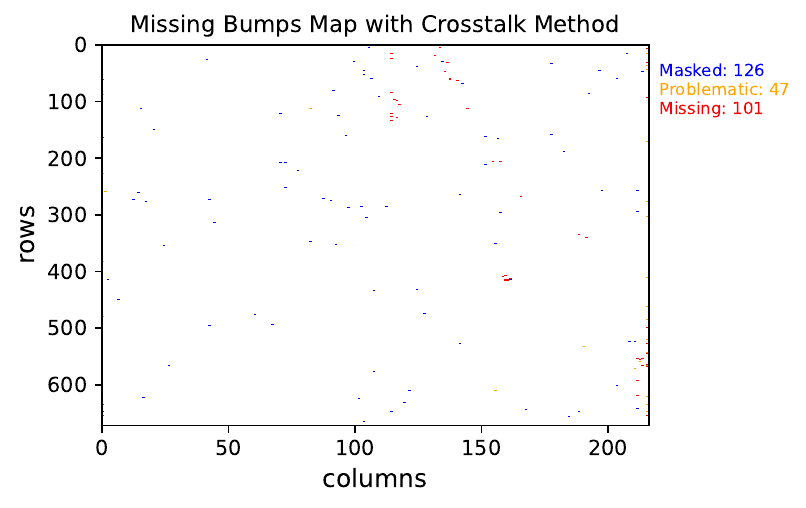} 
		\caption{\label{Missingmap_Chip0_crosstalk}}
	\end{subfigure} \quad
	\begin{subfigure}{0.48\textwidth}
		\centering
		\includegraphics[height=5.0cm, width=\textwidth]{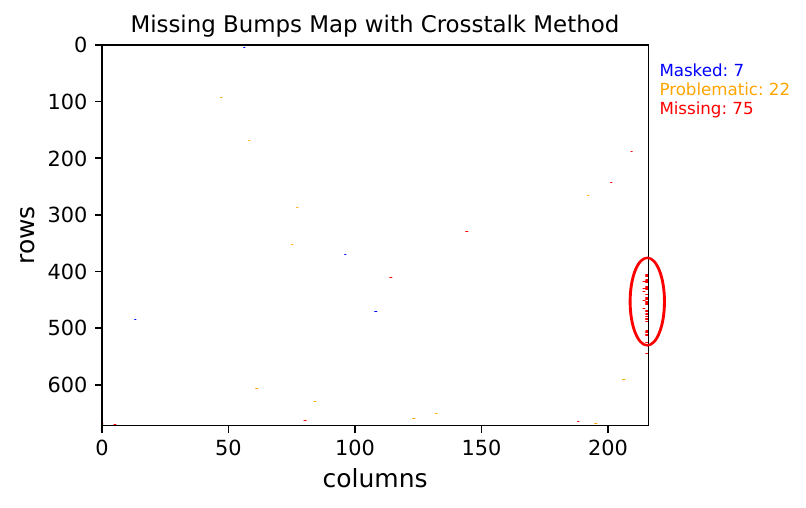}
		\caption{\label{Missingmap_Chip1_crosstalk}}
	\end{subfigure}
	\caption[]{Missing bump classification maps obtained using the crosstalk method for the two bare-modules of the same double module. (a) Chip 0, (b) Chip 1. Pixels are labeled as masked (blue), or missing (red)/ problematic (orange) according to the cross-talk analysis.
		\label{fig:missing_bump_map_crosstalk}}
\end{figure}

Figure \ref{fig:missing_bump_map_crosstalk} shows the bump bond classification results as an example for a double module with 3D sensors, bump bonded by Advafab and assembled at INFN Firenze. The two bare-modules are mounted on a common high-density interconnect (HDI), and pixels are categorized as masked (blue) and problematic (orange) or missing (red) based on their efficiency response to charge injection; all other pixels are considered functional.

\subsection{Reverse/Forward bias method}

The reverse/forward bias method identifies missing bump bonds by analyzing pixel responses under two sensor bias conditions: reverse bias ($-70$ V) and forward bias ($+0.5$ V). In forward bias, a small current is injected into the input of the preamplifier through the sensor. This current modifies the baseline behavior of pixels with intact bump bonds, typically pushing the preamplifier into saturation. In contrast, pixels with missing bumps remain electrically isolated and show little or no change between the two conditions \cite{Starodumov2005}.

To assess bump connectivity, the sensor is tuned to a high threshold (around 6000 e$^-$) and S-curve scans are performed under both bias conditions. A high threshold reduces the influence of noise and enhances sensitivity to baseline shifts caused by the injected current. An S-curve describes the efficiency of a pixel as a function of the injected signal amplitude and has a characteristic sigmoid shape. The inflection point corresponds to the pixel's threshold, while the slope is related to its noise \cite{YeDing2020, CMSPixel2008}. These parameters are extracted for each pixel and compared between the two settings.

\begin{figure}[!htbp]
	\centering
	\begin{subfigure}{0.48\textwidth}
		\centering
		\includegraphics[height=5.0cm, width=\textwidth]{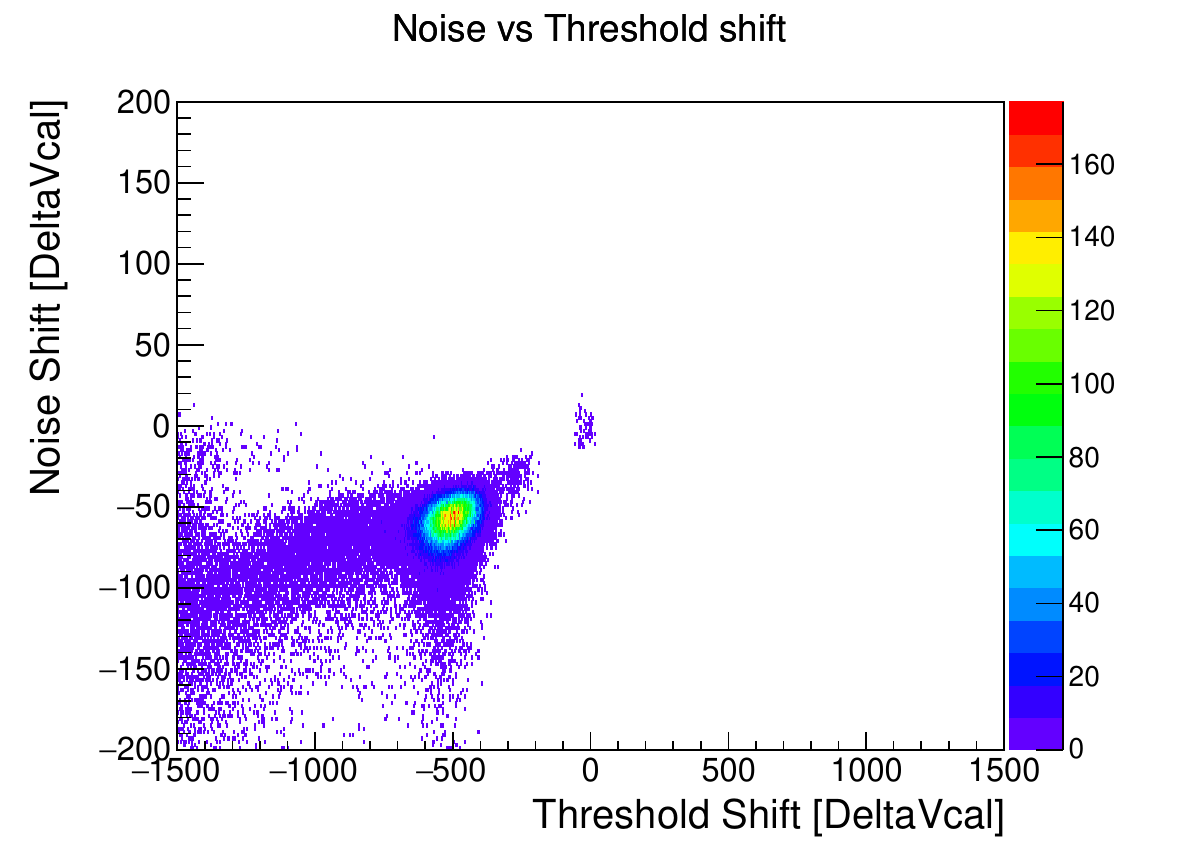} 
		\caption{\label{fig:Chip_0}}
	\end{subfigure} \quad
	\begin{subfigure}{0.48\textwidth}
		\centering
		\includegraphics[height=5.0cm, width=\textwidth]{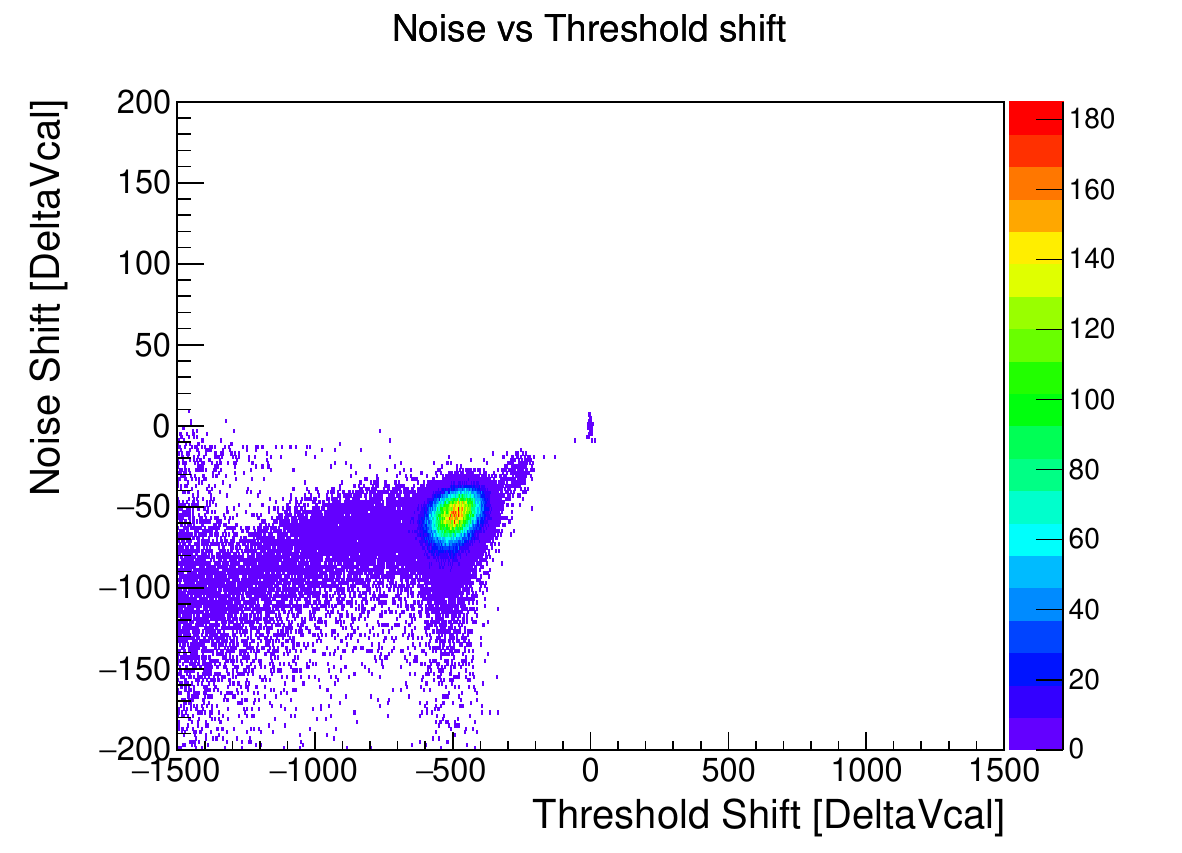}
		\caption{\label{fig:Chip_1}}
	\end{subfigure}
	\caption{2D distributions of threshold and noise shift from the reverse/forward bias method. (a) Chip 0, (b) Chip 1. Connected pixels show large shifts; disconnected ones cluster near the origin.
	}
	\label{fig:2DShift_forwardbias}
\end{figure}

Well-connected pixels typically exhibit a clear shift in threshold and an increase in noise under forward bias. These changes are due to saturation or baseline distortion in the analog front-end. Disconnected pixels, lacking the current injection path, maintain consistent threshold and noise values in both bias modes. The shifts in threshold ($\Delta V_{\text{thr}}$) and noise ($\Delta V_{\text{noise}}$) are used to classify bump status. Figure \ref{fig:2DShift_forwardbias} shows 2D distributions of these values for two bare-modules of a double module. Pixels with both shift values below a predefined threshold are flagged as disconnected. %While these threshold values vary slightly depending on tuning conditions and sensor type, they are typically optimized to balance sensitivity and false positive rates.

\begin{figure}[!htbp]
	\centering
	\begin{subfigure}{0.48\textwidth}
		\centering
		\includegraphics[height=5.0cm, width=\textwidth]{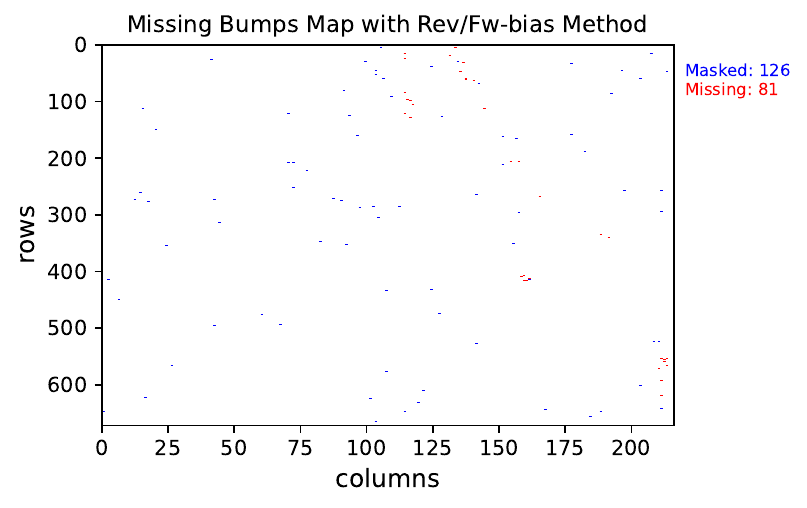}
		\caption{\label{Missingmap_Chip0_fwb}}
	\end{subfigure} \quad
	\begin{subfigure}{0.48\textwidth}
		\centering
		\includegraphics[height=5.0cm, width=\textwidth]{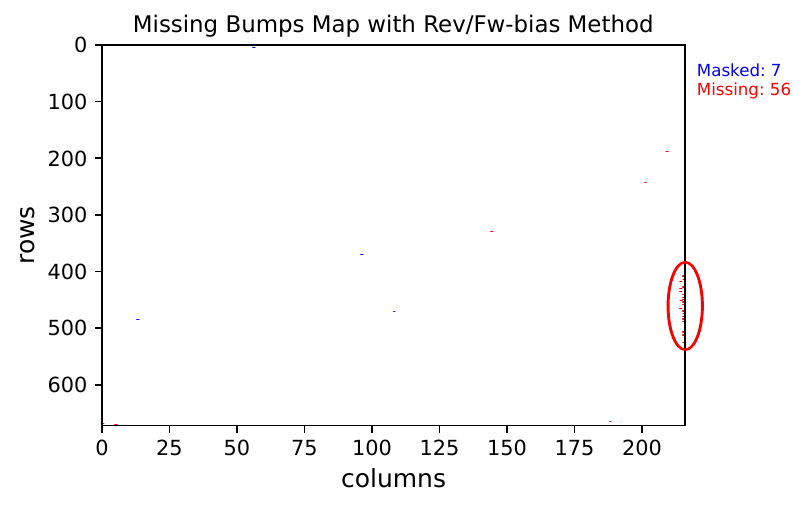}
		\caption{\label{Missingmap_Chip1_fwb}}
	\end{subfigure}
	\caption{Missing bump classification maps using reverse/forward bias. (a) Chip 0, (b) Chip 1. Classification is based on threshold and noise shifts from S-curve comparisons. Pixels are labeled as masked (blue), or missing (red). 
	}
	\label{fig:Missingmap_fwb}
\end{figure}

Reducing the preamplifier bias current was found to enhance the method's sensitivity by narrowing the analog front-end's dynamic range, making it more responsive to baseline shifts from the injected forward current. This effect amplifies threshold and noise changes in pixels with functional bump bonds, particularly in planar sensors where the injected current is smaller due to geometry. In 3D sensors, higher injected currents yield detectable shifts even at nominal bias levels. Figure~\ref{fig:Missingmap_fwb} shows the resulting bump bond classification maps for both bare-modules of a double module, with missing and masked pixels shown in red and blue, respectively.

\subsection{Beta source imaging}

\begin{figure}[!htbp]
	\centering
	\begin{subfigure}{0.48\textwidth}
		\centering
		\includegraphics[height=4.5cm, width=0.8\textwidth]{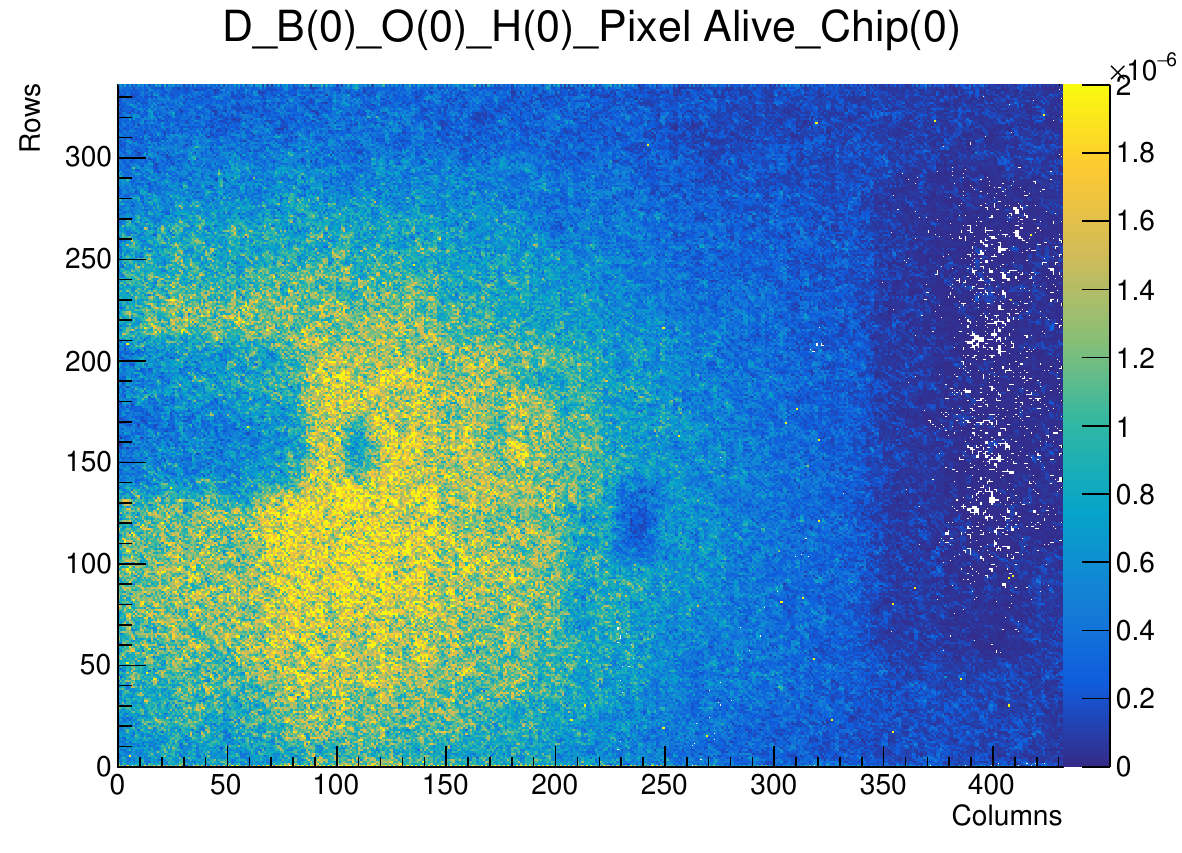}
		\caption{\label{fig:beta_chip0_hist}}
	\end{subfigure} \quad
	\begin{subfigure}{0.48\textwidth}
		\centering
		\includegraphics[height=4.5cm, width=\textwidth]{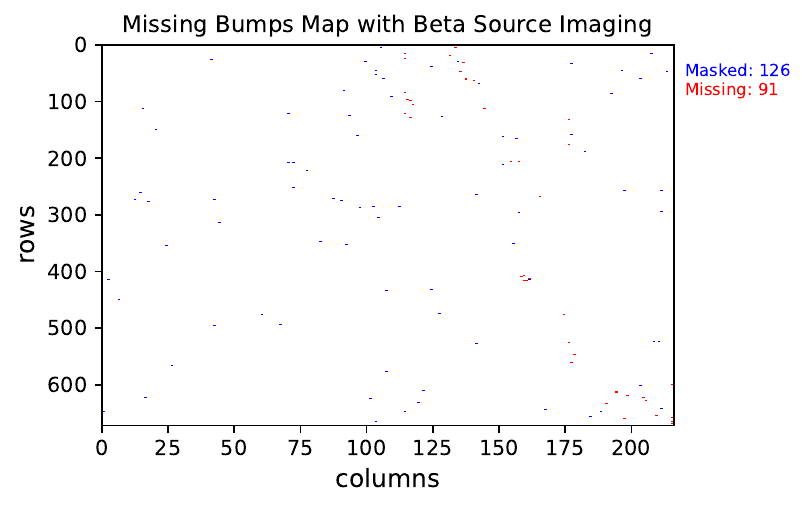}
		\caption{\label{fig:beta_chip0_map}}
	\end{subfigure}
	\caption{Beta source imaging results for Chip 0 of a double module. (a) PixelAlive hit histogram showing the number of hits recorded across the active area. (b) Corresponding bump classification map: missing pixels are shown in red, and masked pixels in blue.}
	\label{fig:beta_chip0}
\end{figure}

\begin{figure}[!htbp]
	\centering
	\begin{subfigure}{0.48\textwidth}
		\centering
		\includegraphics[height=4.5cm, width=0.8\textwidth]{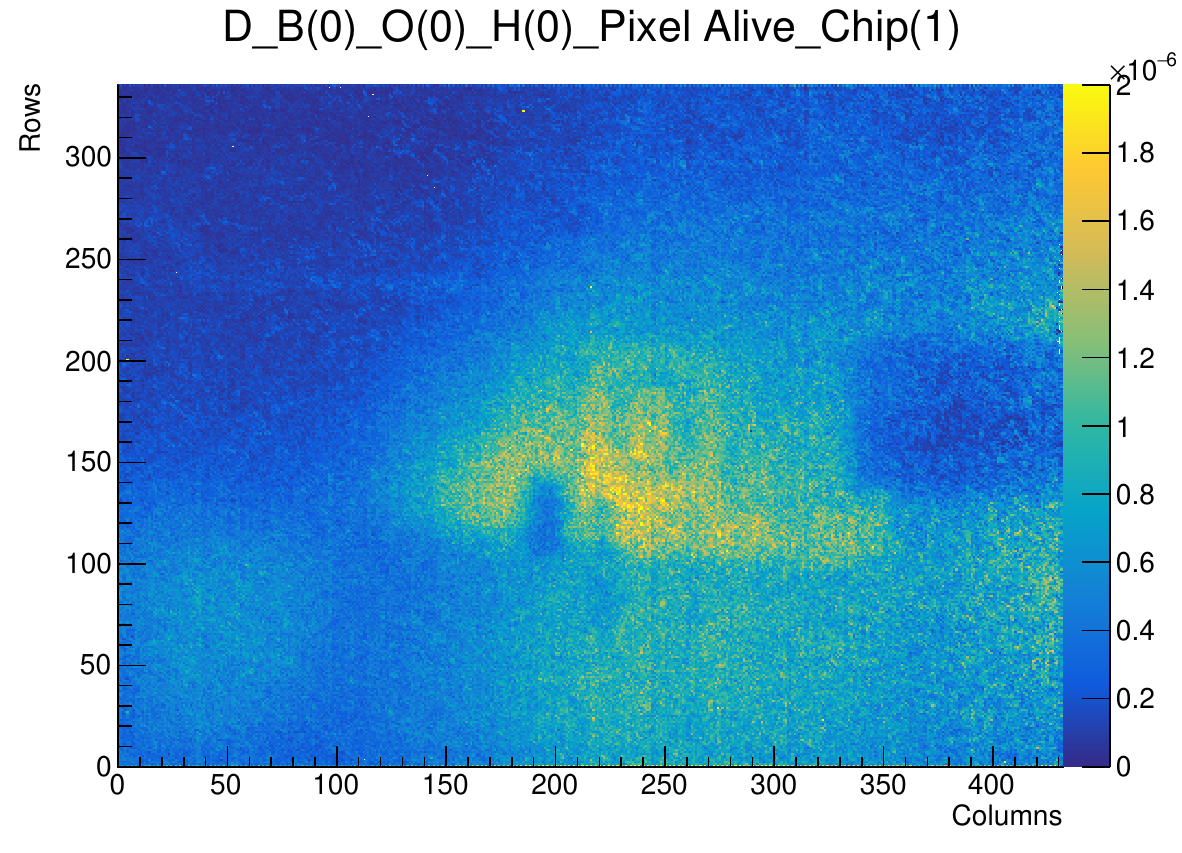}
		\caption{\label{fig:beta_chip1_hist}}
	
	\end{subfigure} \quad
	\begin{subfigure}{0.48\textwidth}
		\centering
		\includegraphics[height=4.5cm, width=\textwidth]{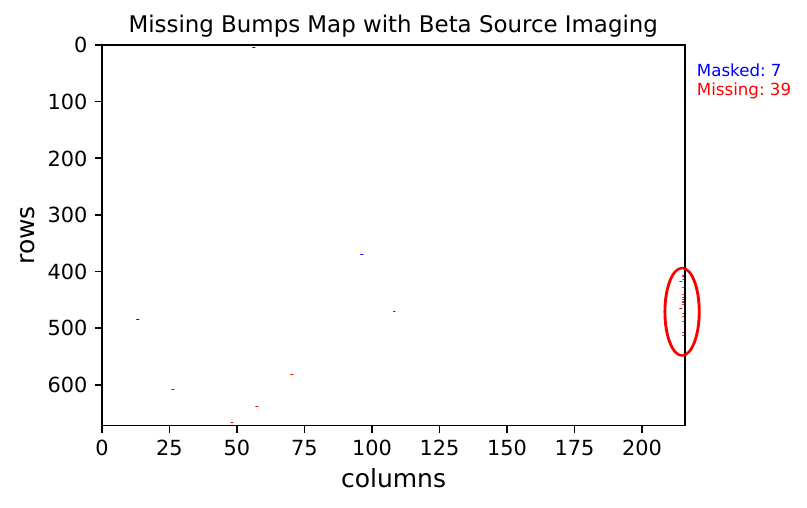}
		\caption{\label{fig:beta_chip1_map}}
	\end{subfigure}
	\caption{Beta source imaging results for Chip 1. (a) PixelAlive hit histogram. (b) Corresponding bump classification map using occupancy thresholds.}
	\label{fig:beta_chip1}
\end{figure}

Beta source imaging provides a direct and reliable method for identifying missing bump bonds. In this approach, the module is uniformly irradiated using a \textsuperscript{90}Sr beta source. Ionization within the silicon sensor generates charge carriers, which are collected by the readout chip through the bump bonds. Prior to irradiation, modules undergo standard calibration procedures, including threshold tuning, noise scans, and pixel masking. Data acquisition is performed with high trigger statistics (typically $\sim10^6$ triggers) to ensure statistical significance. The resulting efficiency maps are then analyzed to identify disconnected pixels, as shown in figures \ref{fig:beta_chip0} and \ref{fig:beta_chip1}.

Pixels located beneath mechanical structures such as the HDI connector may appear inactive because of shadowing effects that reduce particle flux. To avoid misidentifying these pixels as disconnected, they are excluded from the bump classification map based on their known geometric location. Although this method provides a clear physical confirmation of bump connectivity, it is relatively time-consuming and sensitive to non-uniform exposure. It is therefore used primarily to validate suspicious results from electrical tests or for final cross-checks. When combined with other methods, beta imaging offers high-confidence verification of bump bond integrity.

\section{Conclusions}

Ensuring high bump bond yield is essential for the CMS Inner Tracker's performance at the HL-LHC. The methods evaluated in this study probe different physical mechanisms and offer complementary strengths. The crosstalk method is fast, scalable, and applicable to both planar and 3D sensors, making it the default for production, though it may overestimate defects, particularly near sensor edges. The reverse/forward bias method directly assesses electrical continuity and is especially effective for 3D sensors, but it is less sensitive in planar sensors due to lower injection current. Beta source imaging provides high-confidence physical confirmation through charge collection but is time-consuming and more sensitive to shadowing and irradiation non-uniformity. It is therefore mainly used to validate critical modules or clarify ambiguous cases.

Table~\ref{tab:comparison} summarizes the number of missing bumps reported by each method for selected prototype bare-modules. For each method, the table shows the number of missing bumps detected, with the numbers in parentheses indicating how many of the crosstalk-identified missing pixels were also identified by that method. The accompanying percentage quantifies this overlap relative to the total number of missing pixels detected by the crosstalk method. In several cases, the agreement between methods exceeds 60\%, particularly for the reverse/forward bias method. However, lower match rates are also observed--for instance, in FI25 Chip 0--highlighting the distinct sensitivities of each technique. These discrepancies may be due to partially connected bumps that do not produce strong electrical signatures under bias but still respond in beta irradiation via residual capacitive or AC-like coupling.

\begin{table}[h]
	\centering
	\caption{Missing bump counts from different test methods for selected chips. Numbers in parentheses indicate how many of the crosstalk-identified missing pixels were also identified by each method. The percentage represents agreement with the crosstalk method.}
	\label{tab:comparison}
	\begin{small}
		\begin{tabular}{lccc}
			\hline
			Chip & Crosstalk & Reverse/Forward Bias & Beta Imaging \\
			\hline
			FI23 Chip 0 & 101 & 81 (77, 76.2\%) & 91 (79, 78.2\%) \\
			FI23 Chip 1 & 75 & 56 (56, 74.7\%) & 39 (32, 42.7\%) \\
			FI24 Chip 0 & 279 & 227 (195, 69.9\%) & 236 (175, 62.7\%) \\
			FI25 Chip 0 & 71 & 43 (30, 42.3\%) & 17 (15, 21.1\%) \\
			\hline
		\end{tabular}
	\end{small}
\end{table}

The developed methods have been successfully validated on prototype modules and are now integrated into the CMS quality control framework. Their combined use provides complementary sensitivity and redundancy, ensuring that only modules with robust bump bonding are selected for final assembly. This comprehensive strategy is crucial for maintaining the tracking performance and long-term reliability of the CMS Inner Tracker throughout HL-LHC operation.

% Please avoid comments such as "For a review'', "For some examples",
% "and references therein" or move them in the text. In general,
% please leave only references in the bibliography and move all
% accessory text in footnotes.

% Also, please have only one work for each \bibitem.

%\end{thebibliography}

\bibliographystyle{JHEP}
\bibliography{references}

\providecommand{\href}[2]{#2}\begingroup\raggedright\begin{thebibliography}{1}

\bibitem{CMS:2008xjf}
{CMS Collaboration}, \emph{{The CMS experiment at the CERN LHC}},
  \href{https://doi.org/10.1088/1748-0221/3/08/S08004}{\emph{JINST} {\bfseries
  3} (2008) S08004}.

\bibitem{CMS-TDR-014}
{CMS Collaboration}, \emph{{The Phase-2 Upgrade of the CMS Tracker}},  Tech.
  Rep. CERN-LHCC-2017-009, CMS-TDR-014, CERN, Geneva (2017).

\bibitem{Loddo2024}
F.~Loddo et~al., \emph{{RD53} pixel chips for the {ATLAS} and {CMS Phase-2}
  upgrades at {HL-LHC}},
  \href{https://doi.org/10.1016/j.nima.2024.169682}{\emph{Nucl. Instrum. Meth.
  A} {\bfseries 1067} (2024) 169682}.

\bibitem{GRIPPO2022167496}
M.~Grippo, N.~Bartosik, N.~Demaria and F.~Luongo, \emph{Wafer-level testing of
  the readout chip of the cms inner tracker for hl-lhc},
  \href{https://doi.org/https://doi.org/10.1016/j.nima.2022.167496}{\emph{Nuclear
  Instruments and Methods in Physics Research Section A: Accelerators,
  Spectrometers, Detectors and Associated Equipment} {\bfseries 1044} (2022)
  167496}.

\bibitem{RD53CManual}
{\scshape {RD53}} collaboration, \emph{{RD53C Chip Manual}},  Tech. Rep.
  \href{https://cds.cern.ch/record/2890222}{CERN-RD53-PUB-24-001}, CERN (2024).

\bibitem{Starodumov2005}
A.~Starodumov, W.~Erdmann, R.~Horisberger, H.C.~Kästli, D.~Kotlinski,
  U.~Langenegger et~al., \emph{Qualification procedures of the {CMS} pixel
  barrel modules},
  \href{https://doi.org/https://doi.org/10.1016/j.nima.2006.04.087}{\emph{Nucl.
  Instrum. Meth. A} {\bfseries 565} (2006) 67}.

\bibitem{YeDing2020}
Y.~Ding et~al., \emph{The study of calibration for the hybrid pixel detector
  with single photon counting in {HEPS-BPIX}},
  \href{https://doi.org/10.1109/TNS.2020.3006942}{\emph{IEEE Trans. on Nucl.
  Sci.} {\bfseries 67} (2020) 1812}.

\bibitem{CMSPixel2008}
D.~Kotliński, \emph{Status of the {CMS Pixel} detector},
  \href{https://doi.org/10.1088/1748-0221/4/03/P03019}{\emph{JINST} {\bfseries
  4} (2009) P03019}.

\end{thebibliography}\endgroup

\end{document}